\documentclass[pdflatex,sn-mathphys-num]{sn-jnl}

\usepackage{graphicx}%
\usepackage{multirow}%
\usepackage{amsmath,amssymb,amsfonts}%
\usepackage{amsthm}%
\usepackage{mathrsfs}%
\usepackage[title]{appendix}%
\usepackage{xcolor}%
\usepackage{textcomp}%
\usepackage{manyfoot}%
\usepackage{booktabs}%
\usepackage{algorithm}%
\usepackage{algorithmicx}%
\usepackage{algpseudocode}%
\usepackage{listings}%

\theoremstyle{thmstyleone}%
\theoremstyle{thmstyletwo}%

\theoremstyle{thmstylethree}%

\begin{document}

\title[Article Title]{Accelerated Prediction of Surface Stability and Particle Morphology in Ionic Crystals via Electrostatic Screening}


\author[1,2]{\fnm{Sourav } \sur{Baiju}}

\author*[1,2]{\fnm{Payam} \sur{Kaghazchi}}\email{p.kaghazchi@fz-juelich.de}

\affil*[1]{\orgdiv{Materials Synthesis and Processing (IMD-2), 
Institute of Energy Materials and Devices}, \orgname{Forschungszentrum Jülich GmbH}, \orgaddress{\city{Juelich}, \postcode{52428}, \country{Germany}}}

\affil[2]{\orgdiv{ MESA+ Institute for Nanotechnology}, \orgname{University of Twente}, \orgaddress{ \city{Enschede}, \postcode{7500 AE}, \country{Netherlands}}}


\abstract{ This work presents a fast and scalable approach for predicting surface stability and equilibrium crystal morphology in ionic materials using electrostatic analysis. The method constructs stoichiometric slab terminations and evaluates their electrostatic energies, enabling high-throughput screening of surface configurations at a fraction of the cost of conventional approaches. Polar surfaces are identified through surface dipole moment calculations and stabilized via electrostatics-based reconstruction using replica-exchange Monte Carlo simulations. The surface dipole moment further emerges as an effective descriptor to distinguish the behavior of different classes of materials.
By bypassing expensive Density Functional Theory (DFT) calculations, the approach extends naturally to large systems and high-index surfaces that are typically inaccessible to DFT. Electrostatic interactions are shown to capture the dominant trends in relative surface stability across diverse material systems. The method is validated on simple and complex 3D materials as well as 2D layered oxides, where the predicted dominant facets are consistent with reported density functional theory and experimental observations. Importantly, the framework also reveals cases where high-index surfaces play a non-negligible role in the equilibrium morphology.
These results establish electrostatics as a fast and reliable route for high-throughput prediction of surface stability and particle morphology, opening a pathway for accelerated materials discovery and providing a robust starting point for more detailed calculations in complex energy materials.}




\maketitle

\section{Introduction}\label{sec1}

The equilibrium shape of a crystal is determined by the relative stability of its exposed surfaces~\cite{herring1951some,dantzigequilibrium,van1993automated}. Crystal morphology in turn plays a central role in controlling key material properties, including electrochemical performance and catalytic activity~\cite{luo2019morphology,hossain2025exploring}. Accurate surface free energies are therefore essential for predicting equilibrium particle shapes (EPS).
EPS is determined using the Wulff construction, which relates surface free energies to facet surface areas at fixed volume. This approach has been widely applied to battery electrode materials to connect atomistic surface structure to nan-micro particle morphology and electrochemical behavior. For example, first-principles studies of olivine LiFePO$_4$~\cite{wang2007first} and LiMnPO$_4$~\cite{wang2008ab} have shown that a small number of low-index facets, such as $(010)$, $(011)$, and $(201)$, dominate the equilibrium morphology and strongly influence anisotropic ion transport and surface redox processes. Similar strategies have been extended to Na-based cathodes and related polyanion materials, where density-functional-theory (DFT) surface energies for NaFePO$_4$~\cite{whiteside2014particle}, Li$_2$FeSiO$_4$~\cite{hormann2014stability}, and NASICON-type structures~\cite{querel2021role} have been used to construct EPS and analyze interfacial stability and fast-charging behavior.

Despite their success, such first-principles approaches are computationally demanding~\cite{roosen1998wulffman}. Accurate DFT surface free energy calculations require large supercells, consisting of large slab and vacuum regions~\cite{zhang2013equilibrium,reuter2003first}. In principle, all possible  surface orientations and terminations have to be taken into account. These include complex surface structures with many number of atoms.
As a consequence, systematic or high-throughput morphology prediction across materials remains impractical~\cite{yoo2021finite}.
Hence, only a limited number of low-index surfaces are generally investigated.
The surface free energy $\gamma$ of a stoichiometric, non-reconstructed surface is determined by the excess energy of a slab relative to its bulk reference. 

A useful classification of ionic crystal surfaces has been provided by the Tasker et al.~\cite{tasker1979stability}. In this scheme, type~I surfaces are composed of charge-neutral planes and are intrinsically non-polar, type~II surfaces consist of charged planes but remain non-polar due to dipole cancellation, and type~III surfaces are polar and exhibit a net dipole perpendicular to the surface. For polar surfaces, the two slab terminations are inequivalent and the standard surface-energy expression cannot be applied directly. In such cases, charge compensation, stoichiometric reconstructions, and/or dipole corrections must be introduced to eliminate artificial electrostatic fields. 

Electrostatic interactions are the dominant forces in ionic materials. This raises a central question: does electrostatics also determine the relative stability of different surface facets? Short-range bonding and local structural relaxations influence the absolute values of surface energies, but it is not clear whether they control the qualitative ordering of stable and unstable surfaces. When a surface is formed in an ionic crystal, the electrostatic imbalance is not limited to the outermost layer but can extend over several atomic planes. Because of this long-range character, electrostatic contributions may differ strongly between facets and may play a decisive role in their stability. In this work, we examine whether an electrostatics-based description is sufficient to capture the main trends in surface stability and equilibrium morphology of ionic materials.
This motivates the development of a simplified, electrostatics-based approach for estimating relative surface free energies. By considering nominal ionic charges and their geometric arrangement, such a model can capture dominant stability trends among stoichiometric surface terminations at a fraction of the computational cost of full electronic-structure calculations. Moreover, the efficiency of this approach enables systematic screening of  surfaces and their numerous different terminations, which would otherwise be prohibitive using DFT, in particular for high index ones.

In this work, we assess the capabilities and limitations of such an electrostatic surface-energy model by comparing its predictions with available DFT calculations and experimental observations, focusing on relative facet stability and equilibrium particle morphologies. 
We apply this framework to several representative classes of battery materials. We begin with two simple and well-characterized 3D binary compounds, the sulfide Li$_2$S and the oxide TiO$_2$, which serve as benchmark systems with established reference morphologies. We then extend the analysis to more complex materials, including 2D layered LiCoO$_2$ as well as 3D phosphate (LiFePO$_4$) and NASICON-type framework (Na$_3$V$_2$(PO$_4$)$_3$), thereby testing the electrostatic model across a broad chemical and structural landscape.\\ 
\section{Computational Methods}\label{sec2}
\begin{figure*}[htpb]
    \centering
    \includegraphics[width=\textwidth]{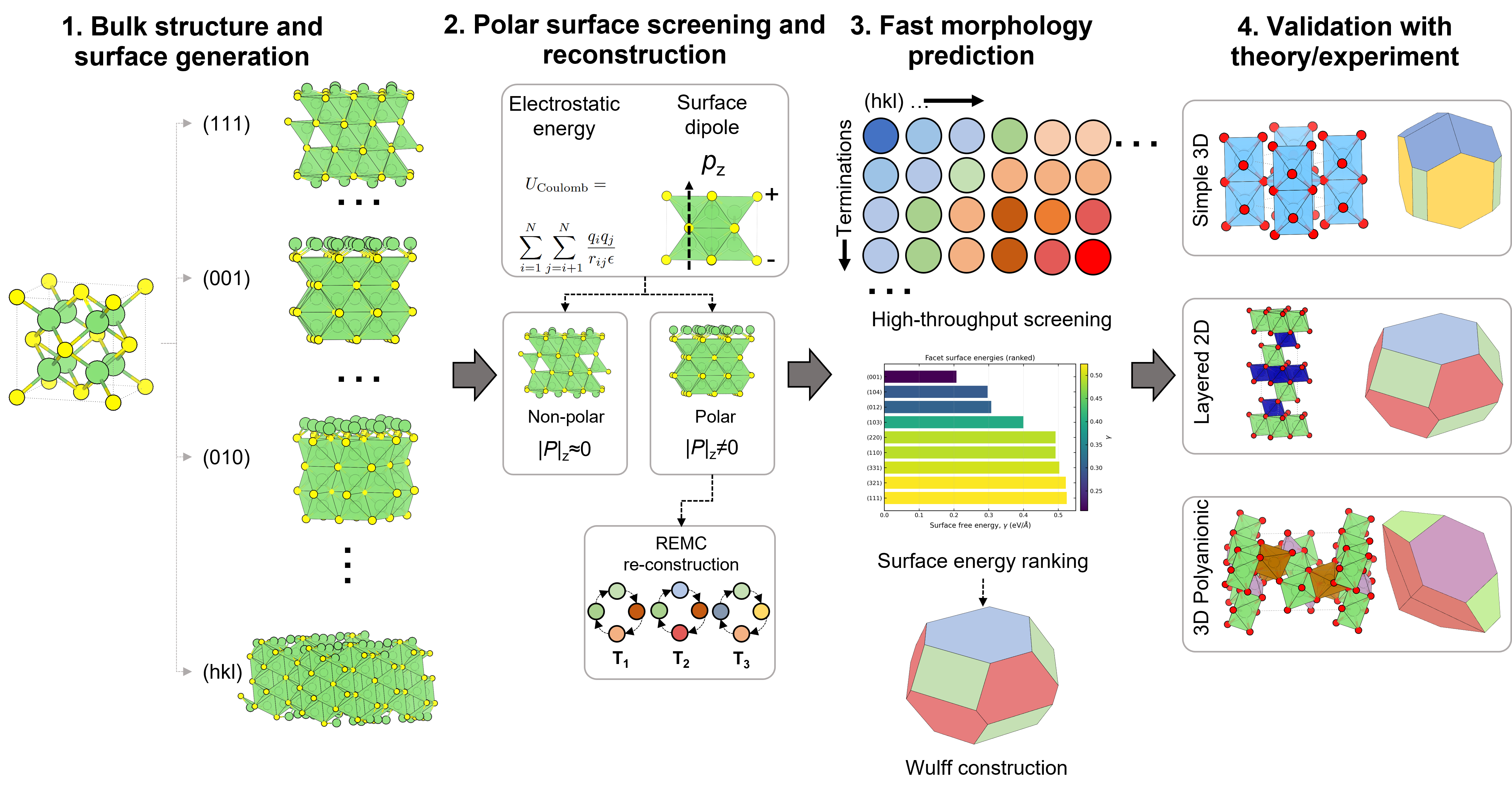}
    \caption{Workflow for electrostatics-based surface-energy evaluation and morphology prediction. Symmetry-distinct slabs are generated from experimental bulk structures, screened for polarity, and evaluated using Ewald-summed electrostatic energies. The lowest-energy termination for each facet family is used in a Wulff construction to obtain the equilibrium particle shape.
}
    \label{fig:workflow}
\end{figure*}
The computational methodology employed in this work is summarized in Fig.~\ref{fig:workflow} and consists of four main steps. First, the bulk crystal structure is used to generate all relevant surface orientations and terminations. Second, polar surfaces are identified and surface reconstruction is introduced to those which can be potentially stabilised. Third, the surface free energy is  estimated for each surface termination using a Coulomb-based approach (hereafter called surface energy, $\gamma$ in equation 1). Finally, the equilibrium particle morphology is determined by computed $\gamma$s using the Wulff construction. The details of each step are described in the following subsections.
\subsection{Surface slab construction and electrostatic treatment}
All bulk crystal structures used in this work are obtained from the Inorganic Crystal Structure Database (ICSD) in their experimentally reported unit-cell form and serve as references for slab construction.
Surface slabs are generated by systematic enumeration of crystallographic orientations defined by Miller indices $(hkl)$ within a defined range, $0 \leq h,k,l \leq h_{\mathrm{max}}$. To avoid redundant sampling of crystallographically equivalent planes, Miller indices are first reduced by their greatest common divisor. Full space-group symmetry operations of the bulk crystal are then applied to group equivalent orientations into symmetry-unique facet families, ensuring that each distinct surface orientation is considered only once.
For each symmetry-unique Miller index, all symmetry-distinct stoichiometric surface terminations are constructed using the slab generation formalism implemented in the \textit{pymatgen} library~\cite{Jain2013}. A tolerance parameter of 0.20 is employed to distinguish inequivalent terminations arising from different atomic layer stackings along the surface normal, while symmetry operations of the bulk space group are used to eliminate redundant slabs related by translational or point-group symmetry. Slab generation and symmetry filtering are performed using a combination of the \textit{pymatgen}~\cite{Jain2013} and Atomic Simulation Environment (ASE)~\cite{larsen2017atomic} libraries. 

To assess the sensitivity of the slab enumeration to tolerance parameter, additional tests were performed using a range of tolerance values. The resulting variation in the number of generated terminations for the Li$_2$S (111) surface is summarized in Appendix figure 1.
Each generated slab is subsequently subjected to a polar-surface screening step based on the dipole moment perpendicular to the surface. To evaluate the dipole efficiently for a large number of slab terminations, we employ a charged-layer (Tasker-type) electrostatic model in which the slab is decomposed into a sequence of atomic planes normal to the surface direction. Formal ionic charges are assigned to each chemical species, and the net charge of plane $n$ is obtained by summing the charges of the atoms belonging to that plane, yielding a plane charge $Q_n$ and surface charge density $\sigma_n = Q_n/A$, where $A$ is the surface area.

The surface-normal direction $\hat{z}$ is defined as $\hat{z} \parallel (\mathbf{a} \times \mathbf{b})$, appropriate for slabs periodic in the $ab$ plane. Atomic positions are projected onto $\hat{z}$ and grouped into planes using a distance tolerance $\delta z$ of 0.1$\AA$ (see appendix figure 2), and a layer center $z_n$ is defined for each plane. Within this framework, the dipole moment along the surface normal can be computed either from the plane charges,
\begin{equation}
p_z = \sum_n Q_n (z_n - z_{\mathrm{ref}}),
\label{eq:dipole}
\end{equation}
The dipole moment reported in this work ($p$) corresponds to the out-of-plane component $p_z$, normalized by the surface area of the slab. Accordingly, the reported quantity represents the dipole moment per unit area and has units of $e/\text{\AA}$.
Slabs with a vanishing dipole moment (within a defined numerical tolerance of > $\approx 10^{-15}$) are classified as non-polar, whereas slabs exhibiting a finite $p$ are identified as polar. For polar terminations, charge compensation and surface reconstruction are carried out to generate physically meaningful slab models. These reconstructions are obtained either through exact Coulomb optimization of the available surface sites or via heuristic optimization using replica-exchange Monte Carlo (REMC) sampling, as implemented in the GOAC code \cite{koster2025optimization}. In the latter approach, the configurational space of possible surface atom and vacancy arrangements is systematically sampled to identify energetically favorable configurations.

All slab models are constructed with a vacuum thickness of $100~\mathrm{\AA}$ on each side of the slab to suppress interactions between periodic images along the surface normal. Convergence with respect to vacuum separation was explicitly tested, and the surface energy is found to be converged for vacuum thicknesses exceeding $10~\mathrm{\AA}$ per side for a non-polar stoichiometric slab (see appendix figure 3).

\subsection{Electrostatic surface energy and Wulff construction}
\label{sec:elec_gamma_wulff}

The electrostatic contribution to the surface free energy is evaluated using a Coulomb-only description based on Ewald summation, as implemented in the GOAC code~\cite{koster2025optimization}. For a system of point charges $\{q_i\}$ located at positions $\{\mathbf{r}_i\}$ in a periodically repeated cell of volume $V$, the total electrostatic energy is written as
\begin{equation}
x_{ij}^{\mathrm{total}} = x_{ij}^{\mathrm{real}} + x_{ij}^{\mathrm{recip}} + x_{ij}^{\mathrm{self}},
\end{equation}
where $x_{ij}^{\mathrm{real}}$, $x_{ij}^{\mathrm{recip}}$, $x_{ij}^{\mathrm{self}}$ are the real-space, reciprocal space interactions and self interaction correction respectively. the individual contributions are mentioned in detail in appendix.

For each slab termination, the electrostatic surface energy is then computed as
\begin{equation}
\gamma = \frac{E^{\mathrm{elec}}_{\mathrm{slab}} - N E^{\mathrm{elec}}_{\mathrm{bulk}}}{2A},
\label{eq:gamma_elec}
\end{equation}
where $E^{\mathrm{elec}}_{\mathrm{slab}}$ is the total electrostatic energy of the slab, $E^{\mathrm{elec}}_{\mathrm{bulk}}$ is the electrostatic energy per bulk formula unit, $N$ is the number of bulk formula units contained in the slab, and $A$ is the surface area of one slab face. The factor of two accounts for the two surfaces present in the symmetric slab geometry.

Using the set of calculated electrostatic surface energies, the equilibrium crystal morphology is determined via the Wulff construction~\cite{barmparis2015nanoparticle}. According to Wulff’s theorem, the equilibrium shape of a crystal with fixed volume minimizes the total surface free energy,
\begin{equation}
\min \sum_i \gamma_i A_i,
\label{eq:wulff}
\end{equation}
where $\gamma_i$ and $A_i$ denote the surface energy and surface area of facet $i$, respectively. This procedure yields the equilibrium particle shape corresponding to the electrostatic surface energetics and allows direct comparison with morphologies obtained from first-principles calculations and experimental observations.

\section{Results and Discussion}\label{sec3}
\subsection{Electrostatic Origin of Polar Surface Instability}
\begin{figure*}[htpb]
    \centering
    \includegraphics[width=\linewidth]{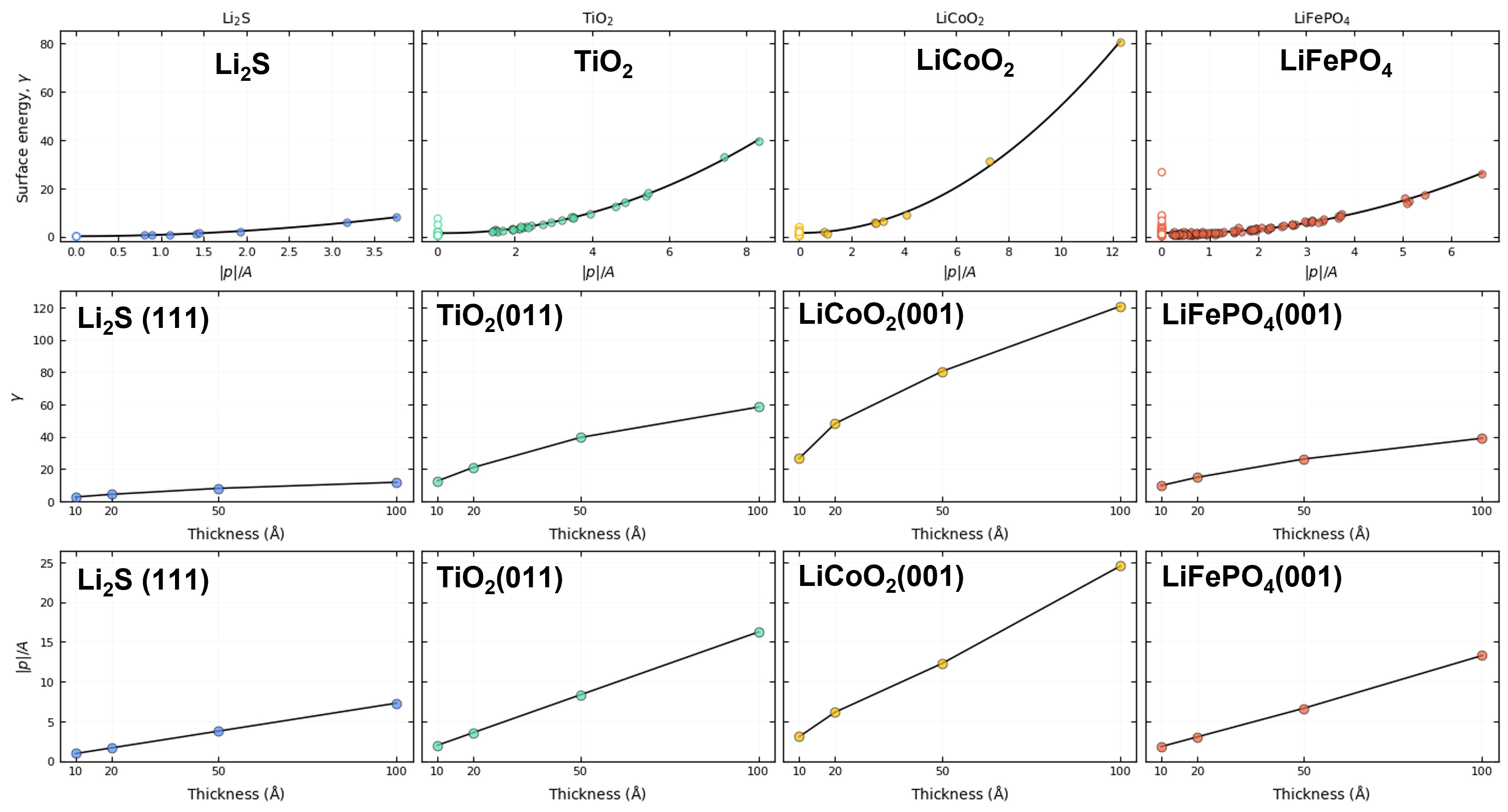}
    \caption{(a–d) Calculated surface free energy as a function of  $\gamma$ for Li$_2$S, TiO$_2$, LiCoO$_2$ (LCO), and LiFePO$_4$ (LFP). The scatter plots show the dependence of $\gamma$ on slab thickness for representative slab terminations, highlighting the approximately quadratic increase in surface energy with thickness for polar surfaces. (e–h) Corresponding slab-thickness dependence of the surface energy $\gamma$ and dipole moment $p$ for the slab terminations exhibiting the largest $\gamma$ and $p$ values in panels (a–d) for each composition.}
    \label{fig:polar-analysis}
\end{figure*}
Figure~\ref{fig:polar-analysis} summarizes the electrostatic surface energies obtained from Ewald summation for all symmetry-distinct slab terminations of four different compositions for which experimental data exists. The calculations were performed using a fixed slab thickness of 50~\AA\ and a vacuum spacing of 100~\AA. This section serves two purposes: (i) to show that all relevant terminations have been systematically studied, and (ii) to verify that the expected electrostatic physics of polar slabs is correctly reproduced with the charged-layer model.

The top row of Fig.~\ref{fig:polar-analysis} shows the electrostatic surface energy $\gamma$ as a function of the absolute dipole moment $|p|$ for all four systems under consideration. Two distinct regions are observed in each curve.\\
(I) several terminations cluster around $p \approx 0$. These correspond to non-polar surfaces. Although their dipole moment is close to zero, their surface energies are not identical. This spread reflects termination-dependent local electrostatic effects, such as different sequences of charged layers (for example bond breaking effects). Even in the absence of a macroscopic dipole, the cleavage energy is determined by the co-ordination and nearest neighbor environment of surface atoms.\\
(II) polar terminations ($p \neq 0$) exhibit a clear quadratic dependence of $\gamma$ on $|p|$. This behavior follows directly from classical electrostatics. The electrostatic field energy is given by
\begin{equation}
U_{\mathrm{field}} = \frac{1}{2} \int \varepsilon E^2 \, dV,
\end{equation}
which is quadratic in the electric field $E$. In a slab geometry, the  electric field generated by a polar stacking scales linearly with the dipole moment per area. For a supercell of height $L_z$ and surface area $A$, we can write 
\begin{equation}
E \propto \frac{p}{A L_z}.
\end{equation}
Substituting this scaling into the field-energy expression yields an electrostatic contribution to the surface energy of the form
\begin{equation}
\gamma_{\mathrm{elec}}(p) \approx \gamma_0 + C p^2,
\end{equation}
where $\gamma_0$ represents the energy cost of bond breaking at the surface, and $C$ is a chemical composition- and geometry-dependent coefficient.
For polar surfaces, surface free energy dependence on dipole moment is found to be very large. In particular for materials with ions of larger charges (TiO$_2$ vs Li$_2$S)and 2D like structure (LCO vs TiO$_2$) $\gamma$ increases more strongly with $|p|$
The clear distinction between surfaces with $p \approx 0$ and $p \neq 0$ demonstrates that the dominant energy cost for polar terminations arises from the macroscopic electrostatic field (caused by dipole), while the spread at $p = 0$ originates from local termination-specific effects.
The bottom row of Fig.~\ref{fig:polar-analysis} investigates the slab thickness dependence for the termination with the largest $|p|$ (and largest $\gamma$) in each material class (from the top row of figure \ref{fig:polar-analysis}). The vacuum thickness is kept fixed at 100~\AA, while the slab thickness is varied from 10 to 100~\AA.
Both the dipole moment and the electrostatic surface energy increase approximately linearly with slab thickness. Within the charged-layer model, each additional repeat unit contributes a nearly constant dipole increment. Under fixed vacuum conditions, the Electrostatic energy cost correspondingly increases with thickness.
Importantly, the slopes differ systematically across materials. The increase is weakest for the simple three-dimensional ionic compound Li$_2$S and strongest for the naturally layered material LiCoO$_2$. This trend reflects the intrinsic stacking of charged planes in layered materials, which generates a larger polarization compared to more three-dimensional materials. The observed scaling therefore can be connected to the crystallography of a given material.

Overall, Fig.~\ref{fig:polar-analysis} confirms that the Ewald-based calculations of atomistic supercell  models reproduces the expected macroscopic electrostatic behavior of polar versus non-polar terminations. The quadratic dependence on dipole moment and the systematic thickness scaling show that the dominant contribution to the increase of polar surface energies originates from classical electrostatics rather than from local bonding effects.

\subsection{\label{sec:level7}Electrostatics-Based Surface Morphologies }
\subsubsection{$\mathrm{Li_2S}$}
\begin{figure}[htpb]
    \centering
    \includegraphics[width=\columnwidth]{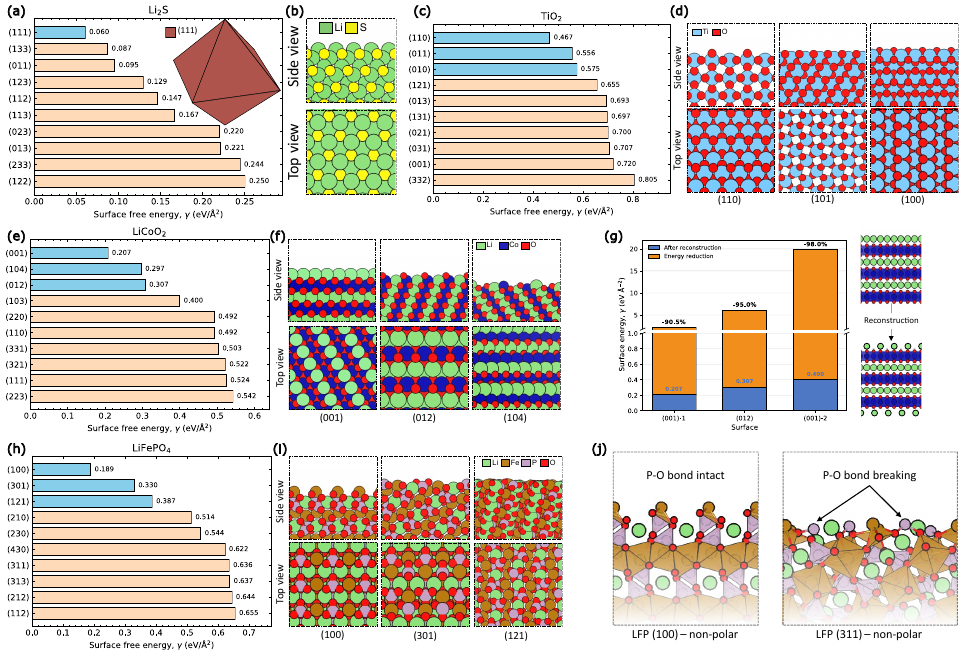}
    \caption{Surface energetics, representative terminations, and reconstruction effects for Li$_2$S, TiO$_2$, LiCoO$_2$ (LCO), and LiFePO$_4$ (LFP). Panels (a,c,e,h) present the ten lowest surface free energies, while panels (b,d,f,i) show the corresponding top and side views of the surface terminations that appear in the Wulff constructions of Li$_2$S, TiO$_2$, LCO, and LFP, respectively. Panel (g) illustrates the effect of surface reconstruction on the surface free energies of the LCO (001) and (012) facets, together with a schematic of the reconstruction scheme applied to the (001) surface. Panel (j) compares the non-polar (100) and (311) terminations of LiFePO$_4$. Cleavage along the (311) plane breaks P--O bonds at the surface, whereas the (100) termination preserves intact PO$_4$ tetrahedra and therefore avoids P--O bond cleavage.}
    \label{fig:gamma-values}
\end{figure}
We begin with Li$_2$S as a test system because of its simple crystal structure and its importance as the final lithiation product in Li--S batteries. Li$_2$S crystallizes in the cubic antifluorite structure ($Fm\bar{3}m$). All surface orientations with $0 \le |h|, |k|, |l| \le 3$ were generated. After symmetry filtering, this resulted in 49 distinguishable slab terminations corresponding to 16 symmetry-distinct facets.
To analyze the electrostatic stability of these surfaces, we evaluated the dipole moments of the slab terminations using the charged-layer model (see Eq.~\ref{eq:dipole}). For the $(111)$ facet, two distinct terminations were identified: one polar and one non-polar. To assess their stability, we studied how the dipole moment and surface energy depend on slab thickness.

Figure~\ref{fig:polar-analysis}e shows the slab-thickness dependence for the polar $(111)$ termination. The dipole moment increases strongly and approximately linearly with slab thickness. The surface energy also increases steadily with thickness, showing that the polar slab becomes less stable. In contrast, the non-polar $(111)$ surface has an almost zero dipole moment for all thicknesses, and its surface energy stays very small and nearly constant at about $\gamma \approx 0.06~\mathrm{eV/\AA^2}$. This comparison shows that the surface energy is strongly related to the calculated dipole.
The broader relationship between dipole moment and surface energy is illustrated in the corresponding scatter plot (Fig.~\ref{fig:polar-analysis}). Non-polar surfaces lie near $|p| \approx 0$ and have small but visible changes in $\gamma$, while polar surfaces show a clear quadratic relation. 

The electrostatic surface energies were then computed for all facets, as shown in Fig.~\ref{fig:gamma-values} (a-b), and used as input for a Wulff construction. The resulting equilibrium morphology is dominated exclusively by the non-polar $(111)$ facet and exhibits a characteristic two-sided cone-like shape (Fig.~\ref{fig:gamma-values}a). This morphology is consistent with previous DFT-based Wulff constructions \cite{chen2014metalization} and with experimentally observed Li$_2$S particle shapes \cite{tan2017burning}.

\subsubsection{$\mathrm{TiO_2}$}

Rutile TiO$_2$ provides a structurally more complex test case than Li$_2$S. It crystallizes in the tetragonal space group $P4_2/mnm$ and exhibits inequivalent Ti–O bonds. From a practical perspective, TiO$_2$ has been widely investigated as a Li-ion battery anode \cite{liu2016recent}. All surface orientations satisfying $0 \le |h|, |k|, |l| \le 3$ were generated together with all symmetry-distinct surface terminations. After symmetry filtering, this resulted in 33 unique slab terminations.

\begin{figure}[htpb]
    \centering
    \includegraphics[width=\columnwidth]{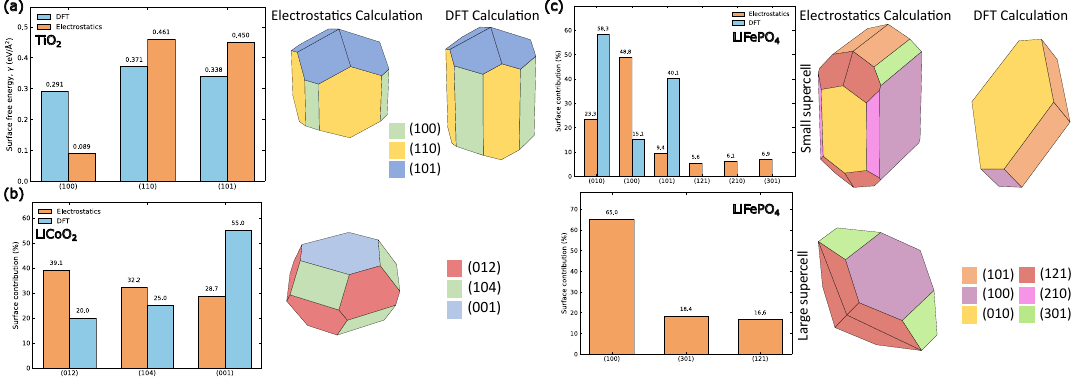}
    \caption{Comparison of equilibrium surface morphologies and facet contributions obtained from electrostatic calculations and density functional theory (DFT). Panels (a) and (b) show the surface contributions (in \%) of the facets present in the equilibrium morphology of TiO$_2$ and LiCoO$_2$ (LCO), respectively, comparing results from DFT and the electrostatic model. The right side of panel (a) shows the equilibrium morphologies of TiO$_2$ predicted from electrostatics and derived from reported DFT surface energies. For LCO, the electrostatically predicted equilibrium morphology is shown in panel (b). Panel (c) compares the equilibrium surface morphologies of LiFePO$_4$ obtained from DFT and electrostatic calculations using two slab models: a small unit cell model and a larger model with a slab thickness of 50~\AA\ and a vacuum size of 100~\AA. For the small model, both DFT and electrostatic approaches give similar morphologies. For the large model, only the electrostatic morphology is shown because the corresponding DFT calculations are computationally too expensive. The associated facet contributions are shown for the small model (top), where DFT and electrostatic results are compared, and for the large model (bottom), where only electrostatic results are available.
}
    \label{fig:wulff-shapes}
\end{figure}

The electrostatic dipole moments of all slab terminations were evaluated using the charged-layer model. Many of the generated terminations are polar and have finite dipole moments. Their surface energies follow the same trend as before: surfaces with small or zero dipole moments have the lowest energies, while strongly polar surfaces have higher energies.
Very small residual dipole values on the order of $10^{-16}$ are observed for some nominally non-polar terminations. These arise from numerical noise and from the finite tolerance used when grouping atomic layers in the charged-layer model. They do not represent physical polarity and have no measurable impact on the computed surface energies.
Using the electrostatic surface energies of all terminations, a Wulff construction was performed (Fig.~\ref{fig:wulff-shapes}a). The resulting equilibrium morphology contains only three dominant facets, namely $(110)$, $(101)$, and $(100)$. This set agrees with DFT-based reference morphologies reported by Jiang \textit{et al.} \cite{jiang2018first}. A comparison with DFT surface energies (Appendix Figure 4) shows that the electrostatic model preserves the same energetic ordering of facets. Minor differences in relative facet area fractions are observed, most notably a reduced contribution of the $(100)$ facet in the electrostatic Wulff shape. These differences likely come from structural relaxation and other interactions not captured in the Coulomb model, but they do not change the dominant facets.

We examined the thickness dependence of the most polar $(011)$ termination. this termination exhibits the largest dipole moment and surface energy among all generated slabs. The dipole moment and surface energy both increase with slab thickness, with a steeper slope than in Li$_2$S, indicating stronger electrostatic effects.
In contrast, non-polar terminations maintain near-zero dipole moments and show no significant thickness dependence in surface energy. 
Overall, the TiO$_2$ results support the same conclusion as for Li$_2$S: surface stability is linked to the dipole. Even in a more complex material with lower symmetry, it remains a reliable way to describe surface stability.

\subsubsection{$\mathrm{LiCoO_2}$ (LCO)}

Layered LiCoO$_2$, a prototypical Li-ion battery cathode, represents a structurally different test case for the electrostatic model than Li$_2$S and TiO$_2$. Unlike the previous systems, LiCoO$_2$ intrinsically exhibits polar surface terminations due to its layered crystal structure. It adopts the O3-type layered structure (space group $R\bar{3}m$), consisting of alternating cation-rich (Li/Co) layers and anion-rich (O) layers stacked along the crystallographic $c$ direction. Cleavage along certain low-index orientations therefore produces asymmetric surface terminations and an intrinsic large electrostatic dipole normal to the slab.

We first focus on the experimentally relevant polar orientations, namely the $(001)$ and $(012)$ surfaces. We explicitly analyze the O-terminated $(001)$ surface, which exhibits the largest dipole moment among the generated terminations. Figure~\ref{fig:polar-analysis}g shows the slab-thickness dependence of the dipole moment and electrostatic surface energy for this termination. As the slab thickness increases from $10$ to $100~\AA$, the dipole moment increases almost linearly. Compared to Li$_2$S and TiO$_2$, the slope of this increase is significantly larger, indicating a stronger accumulation of electrostatic polarization in LiCoO$_2$.
The slab supercell can be viewed as a stack of alternating charged planes. As more layers are added, the charges do not fully cancel, leading to a steady increase in the dipole moment, which is stronger in layered materials like LiCoO$_2$ than in three-dimensional structures. As a result, the electrostatic surface energy of the unreconstructed polar $(001)$ termination increases with slab thickness and does not converge within the investigated range (see Appendix Figure 5). This behavior is characteristic of a Tasker Type~III polar surface. A similar trend is observed for the $(012)$ orientation.
The origin of this instability can be understood from the cleavage geometry. The largest dipole moments occur when the cleavage plane cuts through transition-metal–oxygen octahedra, creating surface layers with large charges. In contrast, Li-terminated surfaces carry smaller charges and show more moderate dipole buildup. A detailed layer-resolved analysis for the two symmetry-distinct $(001)$ terminations is given in Appendix figure 6.

Among all surfaces generated within $0 \le |h|, |k|, |l| \le 4$, the $(001)$ and $(012)$ orientations are the most relevant low-index polar surfaces. Other polar surfaces are either high-index with negligible contributions to the morphology or have much higher energies. While all polar surfaces can in principle be reconstructed, we focus here on $(001)$ and $(012)$ as the main cases.
To obtain physically meaningful surface energies suitable for morphology prediction, polar surface reconstructions were performed following the compensation scheme proposed by Kramer \textit{et al.}~\cite{kramer2009tailoring}. For the $(001)$ orientation, both cation-rich and anion-rich terminations were reconstructed into Li$^+$-rich surfaces by introducing half occupancy in the outermost layers on both sides of the slab(see figure \ref{fig:gamma-values}g). This procedure preserves overall stoichiometry while eliminating the macroscopic dipole moment. An analogous compensation strategy was applied to the $(012)$ surface, resulting in O-rich reconstructed terminations. These reconstructions were generated using a heuristic optimization approach based on replica-exchange Monte Carlo (REMC) as implemented in the GOAC framework.

The effect of reconstruction on the electrostatic surface energy is shown in Fig.~\ref{fig:gamma-values}g, where the $(001)$ and $(012)$ terminations are compared before and after reconstruction. Before reconstruction, the polar surfaces have very large surface energies due to dipole buildup. After reconstruction, the surface energies drop by more than an order of magnitude, showing that the high energies mainly come from electrostatic effects.
After reconstruction, the polar surfaces no longer show strong thickness dependence, indicating that the dipole has been removed. The resulting surface energies are therefore suitable for morphology prediction. Together with the non-polar terminations, these surfaces were used in the electrostatics-based Wulff construction. The predicted morphology (Fig.~\ref{fig:wulff-shapes}b) agrees well with DFT results~\cite{kramer2009tailoring} and is dominated by the $(001)$, $(012)$, and $(104)$ facets, with a significant contribution from the high-index $(104)$ surface. The emergence of the $(104)$ facet highlights the advantage of systematically scanning a broad range of miller indices rather than only simple low index surfaces.

\subsubsection{$\mathrm{LiFePO_4}$ (LFP)}
LiFePO$_4$ (LFP), which crystallizes in the orthorhombic $Pnma$ structure, provides a different test case compared to Li$_2$S, TiO$_2$, and layered LiCoO$_2$. It is widely used as a cathode material in lithium-ion batteries. Unlike simple ionic or layered systems, LFP has a three-dimensional framework made of FeO$_6$ octahedra and PO$_4$ tetrahedra. Because of this more complex structure, it is a good system to study how long-range electrostatics and local interactions together influence surface stability.

We first look at the relation between dipole moment and surface energy for all terminations (Fig.~\ref{fig:polar-analysis}d). As in the previous materials, a clear quadratic trend between $\gamma$ and $|p|$ is observed, showing that strongly polar surfaces have higher energies.
However, the slope of the quadratic trend is intermediate between Li$_2$S and LiCoO$_2$. In particular, the $(001)$ facet, which exhibits the largest dipole moment in LFP, shows a more moderate increase of $\gamma$ and $p$ with slab thickness compared to layered LiCoO$_2$. This behavior reflects the three-dimensional connectivity of the polyanionic framework, which enables partial electrostatic compensation between neighboring layers. For this reason we performed a test by varying the threshold value in the charged layer model (please see method) and identified LFP is the most sensitive to this threshold. As a result, macroscopic dipole buildup is reduced but not completely eliminated.
While the overall quadratic trend holds, the plot shows a clear exception. The high-index $(311)$ surface has a near-zero dipole moment but a very large surface energy of about $27~\mathrm{eV/\AA^2}$. This shows that having no dipole does not always mean the surface is stable.
To elucidate the origin of this behavior, we compare the atomistic structure of the stable non-polar $(100)$ surface with the non-polar but unstable $(311)$ termination (Fig.~\ref{fig:gamma-values}j). The difference comes from how the surface is cut. The $(311)$ plane breaks P–O bonds in the PO$_4$ units, which costs a lot of energy even though the dipole is nearly zero. In contrast, the stable $(100)$ surface keeps the PO$_4$ units intact and avoids this large energy cost. This comparison shows that, in polyanionic frameworks, keeping the tetrahedral units intact is important for surface stability. Although our electrostatic model does not include all bonding effects, it still agrees well with DFT in predicting the relative surface stability and the Wulff shape.

For LiFePO$_4$, the electrostatic surface energies show a clear sensitivity to the supercell size. We generated all symmetry-allowed surface orientations with $0 \leq h,k,l \leq 4$, including all distinct surface terminations, and used these to construct the equilibrium morphology through the Wulff construction.
When a large supercell is used (about $50$~\AA\ thickness and $100$~\AA\ vacuum), the predicted morphology is mainly formed by the $(100)$, $(121)$, and $(301)$ facets (Fig.~\ref{fig:wulff-shapes}c). However, this differs from DFT results, where the dominant surfaces are reported to be $(100)$, $(010)$, and $(101)$ (Fig.~\ref{fig:wulff-shapes}c). Among these, only the $(100)$ facet is common to both approaches. 
To understand this difference, we repeated the calculations using a supercell model similar in size to those used in the DFT study. In this case, the predicted morphology becomes closer to the DFT result, with $(100)$, $(010)$, and $(101)$ as the main facets. Additional facets such as $(121)$, $(210)$, and $(301)$ also appear, but with smaller contributions (Fig.~\ref{fig:wulff-shapes}c). 
This behavior suggests that the predicted morphology of LiFePO$_4$ depends on the supercell size, likely due to its more complex three-dimensional framework. It is also important to note that the $(121)$ facet was not considered in the original DFT study. One advantage of our approach is that it allows large-scale calculations, making it possible to study such complex surfaces. 
To assess the relative surface stabilities, we compared the DFT energies per formula unit for the $(121)$ supercell with those of the $(100)$ and $(010)$ surfaces using slab models comparable to those employed in the reference study. 
The results indicate that the $(121)$ surface is in fact energetically competitive and can even become more stable than the $(100)$ and $(010)$ terminations under comparable supercell sizes (see appendix figure 7). 
This suggests that the absence of the $(121)$ facet in the DFT-predicted morphology likely originates from the limited set of surface orientations considered in that work rather than the difference between electrostatic approach and DFT.
\begin{figure}[htpb]
    \centering
    \includegraphics[width=\columnwidth]{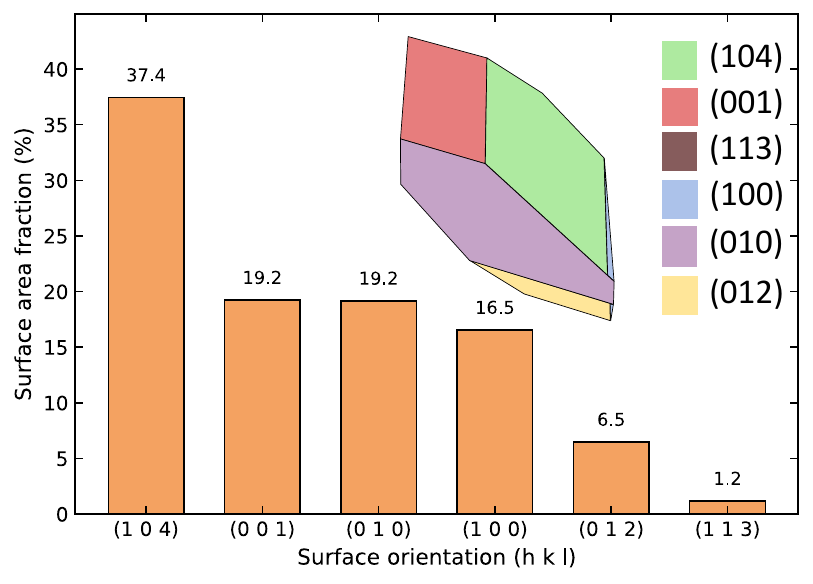}
    \caption{Equilibrium Wulff morphology of Na$_3$V$_2$(PO$_4$)$_3$ (NVP) constructed from electrostatic surface energies. The corresponding surface contributions (in percent) of the facets that appear in the morphology are reported alongside the Wulff shape.}
    \label{fig:nvp}
\end{figure}
Interestingly, when comparing the predicted morphology obtained from the larger electrostatic slab model with experimentally observed LiFePO$_4$ microstructures, a strong qualitative resemblance is observed in tunneling electron microscopy images \cite{jiang2012lifepo4}.
This agreement suggests that the electrostatic model may capture aspects of the equilibrium particle shape that emerge at larger particle sizes, where long-range electrostatic effects become increasingly important.

\subsubsection{Prediction of $\mathrm{Na_3V_2(PO_4)_3}$ (NVP) Particle Morphology}

To extend the electrostatics-based analysis to another three-dimensional polyanionic system, we studied $\mathrm{Na_3V_2(PO_4)_3}$ (NVP), which crystallizes in the rhombohedral $R3c$ structure. NVP is a well-known cathode material for sodium-ion batteries. Using the same workflow as before, we calculated the surface energies for a wide range of Miller indices and constructed the corresponding Wulff shape.
The predicted morphology (Fig. \ref{fig:nvp}) is dominated by the (104) surface, which contributes about 37$\%$ of the total surface area. The (001) and (010) surfaces each contribute around 19$\%$, consistent with the symmetry of the structure. Other surfaces, such as (100), (012), and (113), have much smaller contributions due to their higher surface energies. Overall, the particle shape is a polyhedron with dominant (104) facets and smaller side faces.
This example shows that the electrostatic model can provide reasonable morphology predictions even for complex three-dimensional systems, without relying on DFT calculations or experimental input.

\section{Conclusion}\label{sec13}

In this work, we present a simple and general framework to estimate surface stability and predict equilibrium morphology using electrostatics. The approach combines automated slab generation, identification and reconstruction of polar surfaces, and fast evaluation of electrostatic energies. 
Across all the materials studied, the electrostatic model captures the main energy trends that determine which facets appear in the equilibrium shape. For simple 3D systems such as Li$_2$S and TiO$_2$, the correct low-energy facets are obtained directly from the electrostatic description. For more complex battery materials, including layered LiCoO$_2$ and three-dimensional systems like LiFePO$_4$ and $\mathrm{Na_3V_2(PO_4)_3}$, the method still gives reasonable morphologies. It identifies both low- and high-index facets in line with the structure and known stability trends. Polar surfaces are treated using simple charge-balanced reconstructions, and in three-dimensional frameworks the dipole effects are naturally reduced, leading to stable energy values.
These results show that electrostatics can be used as a simple and low-cost way to estimate surface energies in mostly ionic materials. Moreover, this method provides a fast and scalable way to screen many surfaces. It can be used to guide more detailed calculations and to explore how composition, structure, and morphology are connected. Overall, this framework offers a practical starting point for morphology prediction and large-scale high-throughput computational screening in energy materials research.

\backmatter

\bibliography{refs1.bib}

\end{document}